\documentclass[fleqn,usenatbib]{mnras}

\usepackage[T1]{fontenc}

\DeclareRobustCommand{\VAN}[3]{#2}
\let\VANthebibliography\thebibliography
\def\thebibliography{\DeclareRobustCommand{\VAN}[3]{##3}\VANthebibliography}

\usepackage{graphicx}   
\usepackage{amsmath}    
\usepackage{multirow}

\begin{document}



\title[New Herbig-Haro Objects]{New Herbig-Haro Objects associated with Embedded Sources}

\author[T.A. Movsessian et al.]{
T.A. Movsessian,$^{1}$\thanks{E-mail: tigmov@web.am, tigmag@sci.am }
T.Yu. Magakian,$^{1}$
B. Reipurth$^{2}$
and H.R. Andreasyan$^{1}$
\\
$^{1}$Byurakan Observatory NAS Armenia, Byurakan, Aragatsotn prov., 0213,Armenia\\
$^{2}$Institute for Astronomy, University of Hawaii at Manoa, 640 N. Aohoku Place, Hilo, HI 96720, USA
}

\date{Accepted XXX. Received YYY; in original form ZZZ}

\pubyear{2024}


\label{firstpage}
\pagerange{\pageref{firstpage}--\pageref{lastpage}}
\maketitle

\begin{abstract} 

We continue to present the results of the Byurakan Narrow Band Imaging
Survey (BNBIS). The main goal of this survey is to search for
Herbig-Haro (HH) objects and jets in Galactic dark clouds. In this
work we present the results of the search in the vicinity of
infrared sources that are bright in the WISE survey and embedded in the
dark clouds. The survey is performed with the 1 m Schmidt telescope
of Byurakan Observatory, lately equipped with a new CCD detector,
which allows to obtain one square degree images of the sky in various
filters. Narrow-band filters were used to obtain H$\alpha$ and
[\ion{S}{ii}] images, and a medium-width filter was used for the
continuum imaging.  New HH flows and knots were found near six
embedded IR sources, which constitutes a significant proportion of the
objects observed. At least two of the newly found HH flows (HH~1226
and HH~1227) lie in isolated dark clouds, thus pointing to active star
formation in these regions. Other flows are also located in detached
and dense globules or filaments. The length of the HH~1228 flow is
about 1 pc; it has also a molecular hydrogen counterpart of the same
extension. Coordinates, charts, detailed descriptions and distance
estimates are provided.  The lower limits of bolometric luminosities
of the source stars are typical for low-mass young stellar objects.

\end{abstract}




\section{Introduction}

Since the discovery of Herbig-Haro objects by \citet{Herbig1951} and \citet{Haro1952,Haro1953}, the number of such known objects is now exceeding 1000.\footnote{The catalogue of Herbig-Haro objects is maintained by Bo Reipurth.} 
 These small shock-excited nebulae, visible in low excitation
lines, are the optical manifestation of outflow events near young
stellar objects, in various stages of evolution, mostly at very early
still embedded stages, but also found associated with visible T Tauri 
stars. Very often HHs appear as a string of knots aligned in a
collimated bipolar flow (HH-flow), and sometimes with a bright bow
shock located in a terminal working surface. The spatial extent
covered by HH-flows ranges up to several parsecs (giant
Herbig-Haro flows, \citealt{RBD}). The knots inside such flows
trace mass accretion episodes, and giant HH flows thus provide a
fossil record of the mass loss and accretion history of their sources.

The presence of HH objects is a sign of active star
formation in dark clouds \citep{RB}. 
Surveys for HH flows have led to the discovery of new star formation
regions (SFRs), in particular surveys around small reflection nebulae,
compact stellar groupings, and red nebulous objects in dark clouds
have proven successful. Recently we started a survey (Byurakan Narrow
Band Imaging Survey, or BNBIS) of dark clouds with the 1 m Schmidt
telescope of the Byurakan Observatory, lately equipped with a new CCD
detector, which allows to obtain one square degree images of the sky
in various filters.  The main goal of this survey is the search for HH
objects and jets in Galactic dark clouds. The survey is a continuation
of a search for HH objects that was started with the 2.6 m telescope of
Byurakan observatory more than 20 years ago \citep{MagakianBAO75}, but
with a significantly larger field of view. As a first success the Mon~R1 association, where several new HH objects and outflow systems were
discovered \citep{MMD}, should be mentioned.

In this article we provide data on new HH knots and flows in the
vicinity of bright embedded infrared sources with energy distributions
typical of young stellar objects.

\section{Observations and data reduction}

The images were obtained in 2020-2022 with the 1-m Schmidt telescope
of Byurakan observatory and a 4K$\times$4K Apogee (USA) liquid-cooled
CCD camera as a detector with a pixel size of 0.868\arcsec\ and field
of view of about 1 square degree \citep{Dodo}.  Narrow-band filters
centered on 6560 \AA\ and 6760 \AA, both with a FWHM of 100 \AA, were
used to obtain H$\alpha$ and [\ion{S}{ii}] images, respectively. A
medium-width filter, centered on 7500 \AA\ with a FWHM of 250 \AA, was
used for the continuum imaging.  A log of observations is presented in
Table \ref{log}.

A dithered set of 5 min exposures was obtained in each
filter. Effective exposure time in H$\alpha$ equaled 5100 sec, in
[\ion{S}{ii}] 7200 sec and in the continuum 1800 sec. Images were
reduced in the standard manner using an IDL\footnote{IDL is a trademark of
L3Harris Geospatial Solutions.} package developed by
S.~Dodonov (SAO RAS), which includes bias subtraction, cosmic ray
removal, and flat fielding using a ``superflat field'', constructed from
several images.

The search for HH objects was done with the classic technique,
suggested in 1975 by \citet{VDB}, by comparison of H$\alpha$,
[\ion{S}{ii}] and I-continuum images. It is  based on the characteristic of HH objects strength of [\ion{S}{ii}] emissions, which often is comparable with that of H$\alpha$. To identify probable HH knots we blinked the images, obtained in various wavelengths and also checked all suspicious objects
on the images of the PanSTARRS survey. Over the years, this approach has been
shown to reliably identify HH objects in the overwhelming majority of
cases.
\begin{table*}
\caption{Log of observations}       
\label{log}     
\centering        
\begin{tabular}{l c c c}          
\hline                        
Field & & Obs. date\\ & H$\alpha$ & [\ion{S}{ii}] & Continuum \\    
\hline                     
IRAS 23591+4748 (RNO 150) & 25.08.2020 &  26.08.2020 & 25.08.2020  \\   
IRAS 01166+6635 & 18.12.2020 &  20.13.2020 & 18.12.2020  \\
LkH$\alpha$ 101 & 13.10.2020 & 14.10.2020 & 13.10.2020 \\
2MASS 06590141$-$1159424 (RNO 80) & 11.02.2021 & 05.03.2021 & 11.02.2021\\
IRAS 19219+2300 & 12.09.2021 & 20.06.2022 & 12.09.2021   \\
IRAS 20472+4338 & 18.07.2020 & 18.07.2020 & 18.07.2020 \\
\hline                                       
\end{tabular}
\end{table*}

\section{Results}

\subsection{Sample selection criteria}
In the initial stage of the survey we chose as targets young stellar
objects associated with small reflection nebulae, primarily with a
characteristic cone shape. Later the survey was expanded with a search
in the vicinities of selected far-infrared sources in little-known
small dark clouds and globules. Preliminary results for several fields
were shown during a conference in Byurakan  \citep{movetal2021}.  For the present paper we surveyed embedded IR sources with
characteristic SEDs and specific positions in color-color
diagrams.

A list of the detected outflows and their potential sources is
presented in Table \ref{HHos}. It contains their HH numbers,
coordinates, IRAS or 2MASS names of the probable sources, distances
and position angles of knots from the sources as well as lower limits
to lengths of the flows. In the following we describe results for
each flow separately in order of right ascension.

\begin{table*}
\caption{List of newly discovered HH objects}       
\label{HHos}     
\centering        
\begin{tabular}{c c c c c c c}          
\hline                        
 HH knot & RA (2000.0) & Dec (2000.0) &IR source & r\arcsec & P.A.\degr & length (pc) \\
   & \ \ h \ m \ s & \ \ \degr\ \ \arcmin\ \ \arcsec\ \\    
\hline                     
 HH 1226 (RNO 150) & 00 01 44.1 & +48 06 01 & IRAS 23591+4748 &43 & 11 & 0.08\\
 HH 1227 C & 01 20 02.9 & +66 51 00 & \multirow{3}{*}{$\Biggr\}$IRAS 01166+6635} & 37 &\multirow{3}{*}{$\Biggr\}$ 189} & 0.15\\
  HH 1227 B & 01 20 03.7 & +66 51 27 &  & 9 & & 0.04 \\
  HH 1227 A & 01 20 03.8 & +66 51 33 &  & 4 & & 0.02     \\
 HH 1228 A & 04 30 42.8 & +35 26 16 &  \multirow{5}{*}{$\Biggr\}$WISEA J043041.15+352941.4} & 207 & 175 & 0.53\  \\
 HH 1228 B & 04 30 42.6 & +35 26 06 & & 221 &175 & 0.57 \\
 HH 1228 C & 04 30 42.9 & +35 25 36 & & 247 & 175 & 0.63 \\
 HH 1228 D & 04 30 44.1 & +35 24 50 & & 297 & 172 & 0.76\\
 HH 1228 E & 04 30 48.0 & +35 23 27 & & 385 & 168 & 0.99\\  

 HH 1229 (RNO 80)  & 06 59 00.1 & $-$12 00 15  & 2MASS 06590141$-$1159424 & 38 & 216 & 0.22\\
HH 1230 & 19 24 03.8 & +23 06 21 & IRAS 19219+2300 & 40 & 107 & 0.10 \\
 HH 1231 A & 20 48 56.8 & +43 50 03 & \multirow{2}{*}{$\Bigr\}$IRAS 20472+4338} &33 & \multirow{2}{*}{$\Bigr\}$ 255} & 0.14 \\
 HH 1231 B & 20 48 52.5 & +43 49 57  & & 82 & & 0.36
\end{tabular}
\end{table*}

\subsection{Description of individual objects}

\subsubsection{IRAS~23591+4748 and the RNO~150 field \\ }

IRAS~23591+4748 is a Class I young stellar object with signs of activity
\citep{Connelley_Greene2010,Connelley_Greene2014}. The star is visible at both optical (Gaia DR3~393299436722321664) and 
infrared (2MASS~J00014325+4805189) wavelengths. It is located near the edge of the isolated dark
cloud CB~248 \citep{CB1988}. \citet{CRT2009} found it to be a close
binary with about 1\arcsec\ separation.  Its distance determined in
Gaia DR3 is about $380\pm16$ pc. 

A small and very red nebulous object is located about 43\arcsec\ north
of IRAS~23591+4748 near the center of a dark globule. Almost certainly
this nebulous patch, which is visible even on DSS charts, is the
RNO~150 object, because it matches well the description in the list of
\citet{cohen1980}, but the coordinates provided by Cohen are not
precise. 

\begin{figure}
  \centering
  \includegraphics[width=\columnwidth]{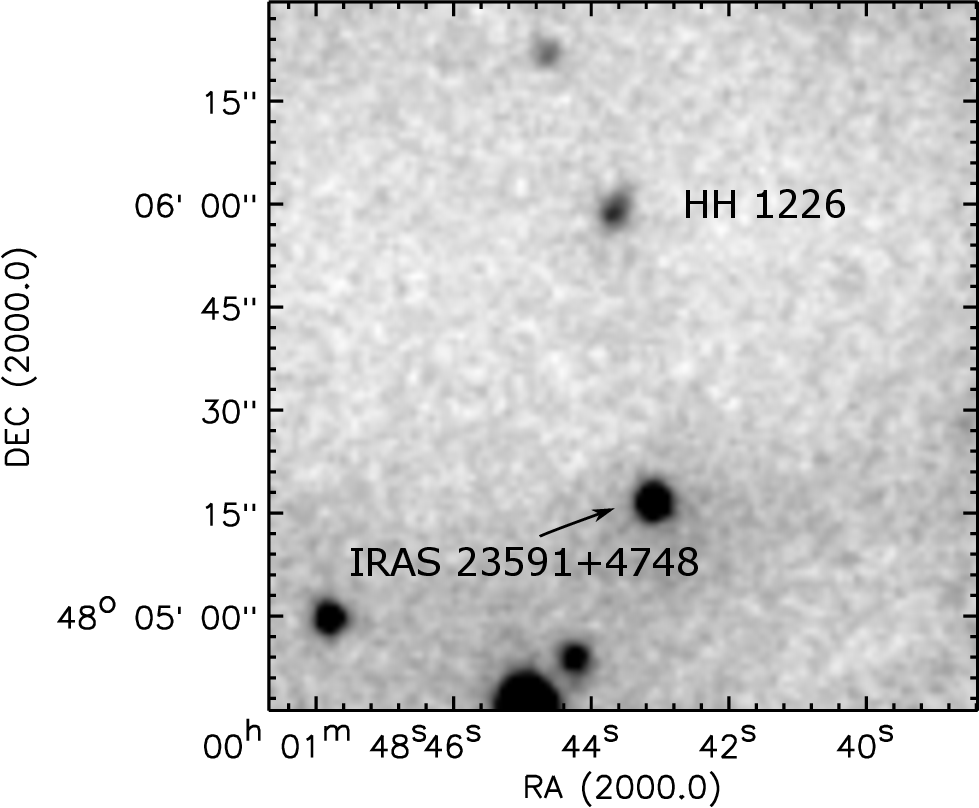}
   \caption{IRAS~23591+4748 and RNO~150 (i.e. HH~1226) as seen in an
   H$\alpha$ image (all images in this paper, unless otherwise
   specified, are obtained with the 1 m Byurakan Schmidt telescope).}
   \label{RNO150}
\end{figure}

Our observations indicate the HH nature of this object, which is
visible both in H$\alpha$ and [\ion{S}{ii}] images while lacking in
continuum emission (Fig.\ref{RNO150}). In both images one can see
a faint short tail, directed to NW, which is distinctly brighter in
H$\alpha$. We here label this object as HH~1226.

It should additionally be noted, that HH~1226 can be seen in the \textit{K}
band image of 2MASS survey as well as in W2 (4.6 $\mu$m) band of
unWISE survey. This indicates the presence of H$_2$ emission in
the object.

\subsubsection{The IRAS~01166+6635 field}

\begin{figure}
  \centering
  \includegraphics[width=23PC]{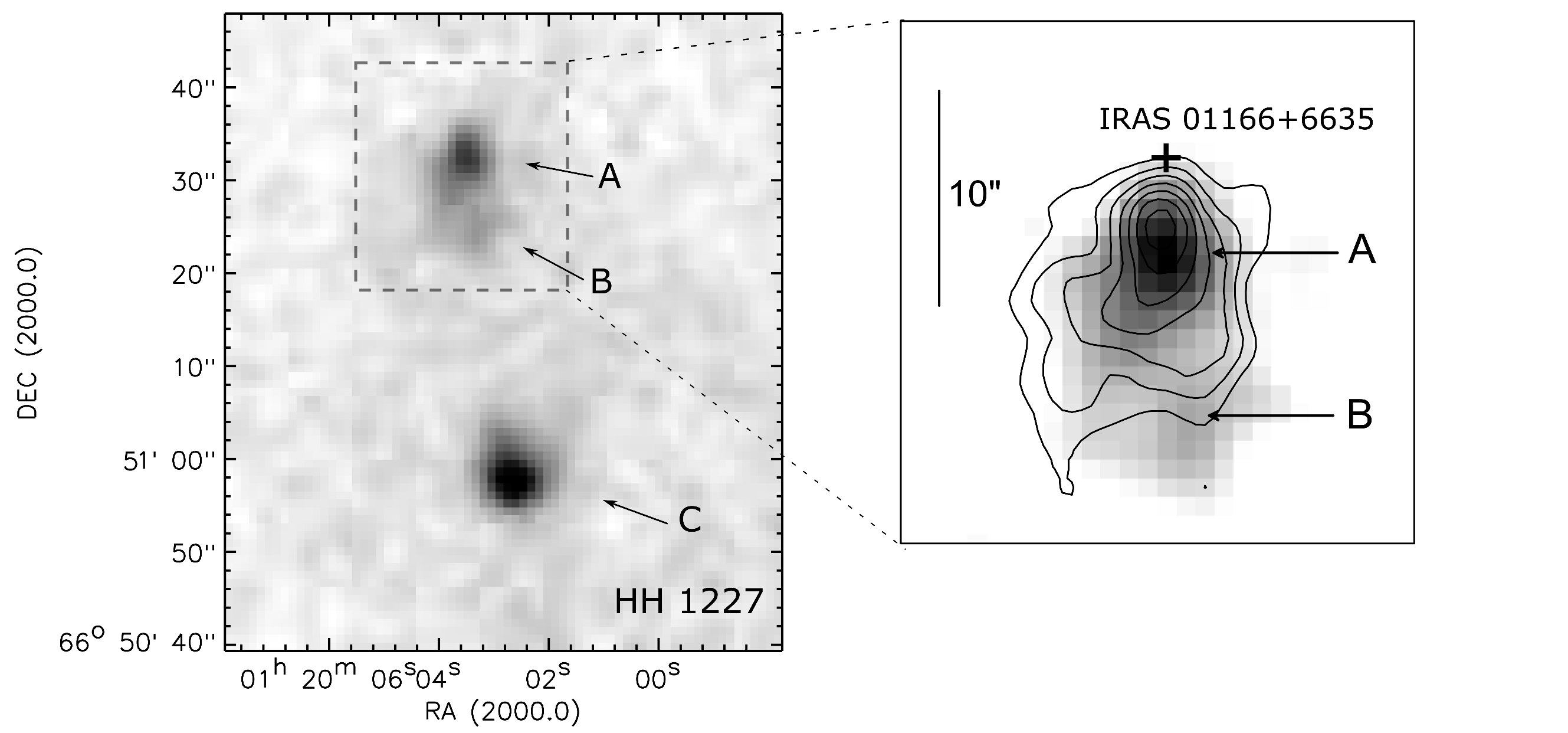}
   \caption{Left: the nebula, connected with IRAS~01166+6635 source and its HH flow (HH~1227 A, B and C); H$\alpha$ + [\ion{S}{ii}] image. Right: an enlarged part of the cometary nebula, as seen in continuum (isolines) and in H$\alpha$ + [\ion{S}{ii}] emission (greyscale).  The spots of enhanced line emission, marked as A and B, do not coincide with the details of the image in continuum.  }
   \label{IRAS01166}
\end{figure}

IRAS~01166+6635 is a very red star visible in PanSTARRS
\textit{i} images.  It is bright at near-infrared  and especially at 
mid-infrared wavelengths (it is listed as 2MASS 01200392+6651358
and WISE~J012003.93+665135.9). The source is associated with a faint
cone-shaped reflection nebula oriented in a N-S direction, 10\arcsec\
in length, which is well seen in our continuum and H$\alpha$
images. It is located near the center of the isolated dark cloud TGU~829 \citep{Dobashi2005} or Dobashi~3782
\citep{Dobashi2011}.  This nebula is also visible in the near-IR range,
with smaller dimensions and oriented to the south-east
\citep{CRT2007}. At a distance of 37\arcsec\ to the south from this
source we found a conspicuous HH object, right on the axis of the
optical reflection nebula. It is elongated in a northern direction for
about 6\arcsec\ and consists of a compact leading knot, seen in
H$\alpha$ and [\ion{S}{ii}], and a diffuse trail, visible mostly in
H$\alpha$. This HH knot was discovered independently in the IGAPS
survey; its appearance and kinematics are described in a recent paper
\citep{IGAPS}. Detailed examination reveals the existence of another small
emission knot, brighter in [\ion{S}{ii}], near the southern edge of
the nebula and about 12\arcsec\ from the central star.   Our [\ion{S}{ii}] images also
show enhanced brightness in the 2\arcsec\ zone just to the south
from the brightness maximum in the reflected light (which definitely contains significant part of the reflected stellar
H$\alpha $ emission) - see Fig.\ref{IRAS01166}, right panel. This might suggest the existence of a jet
near the star, which is not visible in the optical.   We label the separate knots as HH~1227 A, B and C.

As one can see, the IR source (marked by cross in
Fig.\ref{IRAS01166}, right panel) is offset to the north from the tip
of the reflection nebula because of the increased extinction toward
the star. The same effect is seen in the RNO~80 object (see below).  

We did not find any indications of H$_2$ emission from the HH~1227 flow
in the images of 2MASS and unWISE surveys.

In the paper by \citet{IGAPS} a kinematical distance of 240 pc,
estimated from CO and H$_2$O radio observations by \citet{WB89}, is
given for this cloud. On the other hand, the same CO
survey gives for IRAS~01160+6529, which is located only one degree
in angular distance from IRAS~01166+6635, several kinematic estimates,
including 830 and 570 pc. Since the catalog of \citet{Zucker2020}
gives 834$\pm$41 pc for the nearby L~1307 dark cloud, estimated
directly from \textit{Gaia} trigonometric parallaxes, 830 pc seems
a more reasonable value for the distance of IRAS~01166+6635. This value
yields 0.15 pc for the projected length of the HH~1227 flow.

\subsubsection{WISEA J043041.15+352941.4 }

North of the LkH$\alpha$~101 star forming region \citep[see
the review of][]{Andrews_Wolk} there are a number of
nebulous stars, surrounded by wisps of dark matter, which never have
been studied in detail. In particular, no HH objects were found in
this region. Our observations have identified several patches of what
appears to be collisionaly excited emission in the area of the small
reflection nebula around the 12$^{m}$ star Gaia DR3~173369863892701312, separated from LkH$\alpha$~101 by a dark lane,
which is part of LDN~1482.

\begin{figure}
  \centering
  \includegraphics[width=\columnwidth]{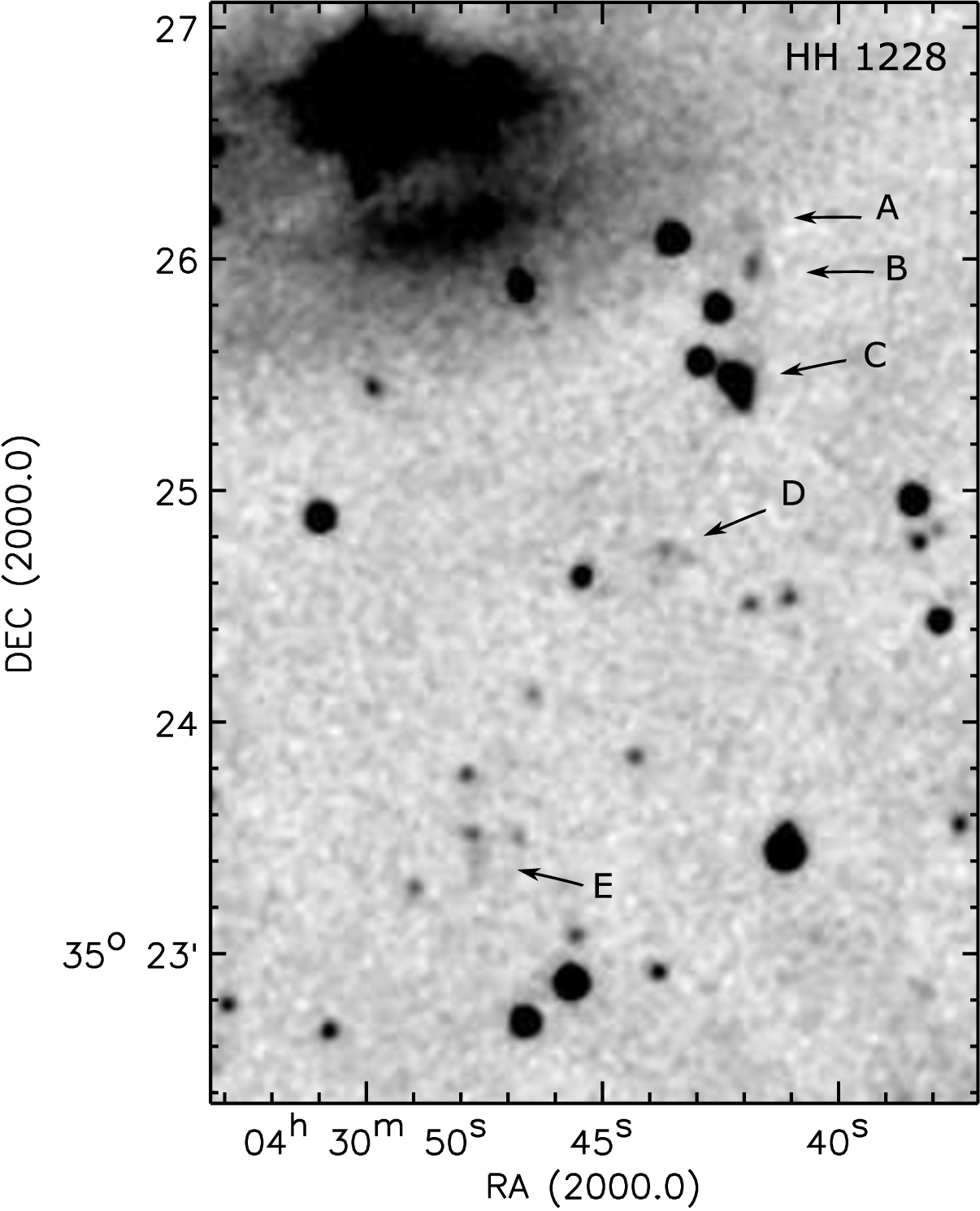}
   \caption{ H$\alpha$ + [SII] image of the HH~1228 flow, located to the north of LkH$\alpha$~101. Individual knots are marked by letters. The probable driving source WISEA~J043041.15+352941.4 is outside of the upper border of this frame. }
   \label{lk101}
\end{figure}

One of these knots (C) is particularly bright (see Fig.\ref{lk101})
and has an oblong shape, oriented perpendicularly to the probable
direction of the flow. It has also a small round appendage. Other four
emission patches are very faint. 

The comparison of our data with a 4.6 $\mu$m image of the unWISE survey confirmed the reality of all
knots as well as the presence of H$_{2}$ emission in them.   There is
no obvious source of this flow (which we have labeled HH~1228) at either
optical or near-IR wavelengths. However, tracing the flow in a
northerly direction, one can find in allWISE images a very red object
WISEA~J043041.15+352941.4, not detected even in the 2MASS survey. This
source lies about 15\arcmin\ to the NNW from LkH$\alpha$~101.

\begin{figure}
  \centering
  \includegraphics[width=\columnwidth]{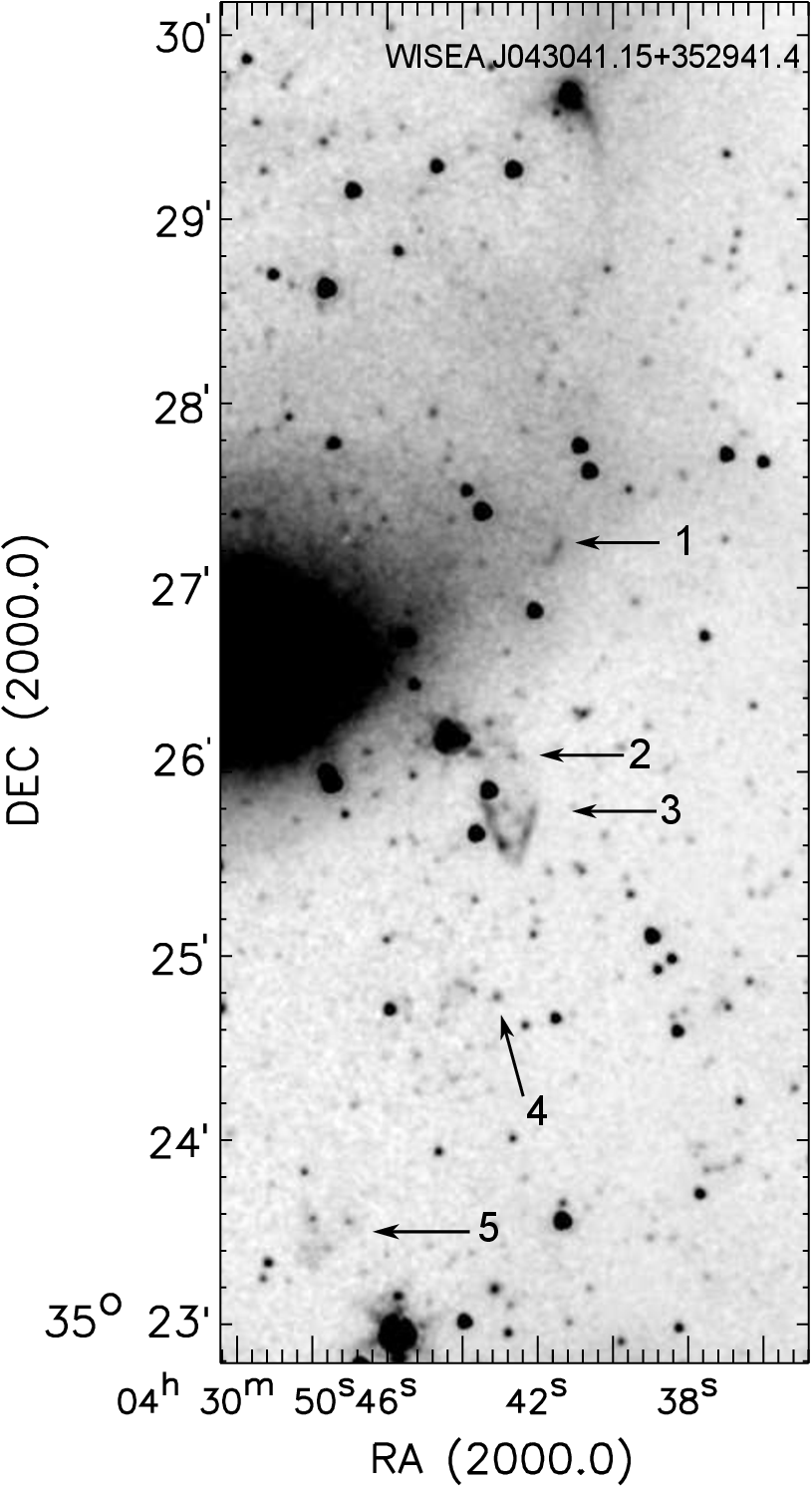}
   \caption{4.5 $\mu$m image of the H$_2$ flow, associated with HH~1228. Individual components are marked by numbers. }
   \label{SEIP}
\end{figure}

Analysing the images of the same field in the Spitzer mid-IR SEIP survey, which have significantly higher resolution, we found that all HH knots of the flow indeed are well visible on the 3.6 and 4.5 $\mu$m images. On Fig.\ref{SEIP}, where the  H$_{2}$ condensations are marked by numbers, object ``2'' corresponds to HH~1228~A+B, object ``3'' - to HH~1228~C, ``4'' - to HH~1228~D and ``5'' - to HH~1228~E. Moreover, on the place of the knot HH~1228~C in H$_{2}$ emission we see a well-defined bow-shock, oriented to the south, while the emission near the knot E is more like a bow-shock with the axis pointing to the south-east. Besides, at a distance of 2.5\arcmin \ to the south from WISEA~J043041.15+352941.4 on the SEIP images we found one more arcuate knot with the same orientation, which has no optical counterpart (``1'' in Fig.\ref{SEIP}).  The images clearly show the IR reflection cometary nebula, illuminated by the WISEA~J043041.15+352941.4 source, the axis of which is directed toward the HH~1228 flow and its molecular analogue (see Fig.\ref{SEIP}, upper side). Thus, we have no reason to doubt
that this IR object indeed is the
driving source of the HH~1228 flow. If this
assumption is correct, then this flow starts near the source at
a position angle of 175\degr\ but afterwards, near knot C, turns
slightly in a direction with position angle 156\degr.

WISEA J043041.15+352941.4 is included in the survey of young stellar
objects in the Gould Belt by \citet{SST} under the designation \mbox{SSTgbs
J0430411+352941}. According to the results of their photometry and
modelling after a correction for A$_{V}$=7.5, this extremely red
object has a very low T$_{dust} = 53$ K, while its bolometric
luminosity is estimated as 2.5 L$_{\sun}$. 

The distance to the LkH$\alpha$~101 star forming region has been quite
controversial \citep{Andrews_Wolk}, and even most recent estimates,
based on the trigonometrical parallaxes of Gaia DR3, somewhat
vary. According to the catalog of \citet{BJ2021}, LkH$\alpha$~101
itself has a distance about $\approx 620 \pm 40$ pc (but with
significant measurement error: RUWE=2.36), while \citet{dzib}
estimated a distance of $ 535 \pm 29$ pc for one of the members of the
LkH$\alpha$~101 cluster by VLBA radioastrometry. To understand if the
nebulae and dark lanes, located in the northern direction from this
star forming region, have the same distance, we checked the Gaia
parallaxes of two bright stars illuminating nebulae to the east and to
the west of the HH 1228 flow: the above mentioned TYC 2381-482-1 (Gaia DR3
173369863892701312) and 2MASS J04301644+3525217 (Gaia DR3
173380579834746112). The distance of both stars in the \citet{BJ2021}
catalog is very near to 530 pc, in excellent agreement with the
radioastrometric distance of the LkH$\alpha$~101 cluster. Thus, we
adopt 530 pc as the distance for the HH~1228 flow. If WISEA
J043041.15+352941.4 is indeed the source of this outflow, then its
projected length, measured along the flow, will be about
1.0 pc, which makes it the most extended flow in the present sample.

\subsubsection{RNO~80 (2MASS~06590141$-$1159424)}

RNO~80 is a faint, nebulous object in the dark cloud Dobashi~5068. It was discovered by \citet{cohen1980}, who described it
as a group of faint stars in a common nebula, but no stars can be seen
in the optical on PanSTARRS images. The nebula has the appearance of a
small cylindrical lobe with a rounded bright edge. The exciting star
starts to be visible only in the \textit{H} band image of the 2MASS
survey; in \textit{K} band it already becomes prominent. As in many
similar cases, the infrared star is embedded and offset from the
optical nebula which likely represents an illuminated cavity, created by a molecular
outflow.

\begin{figure}
  \centering
  \includegraphics[width=\columnwidth]{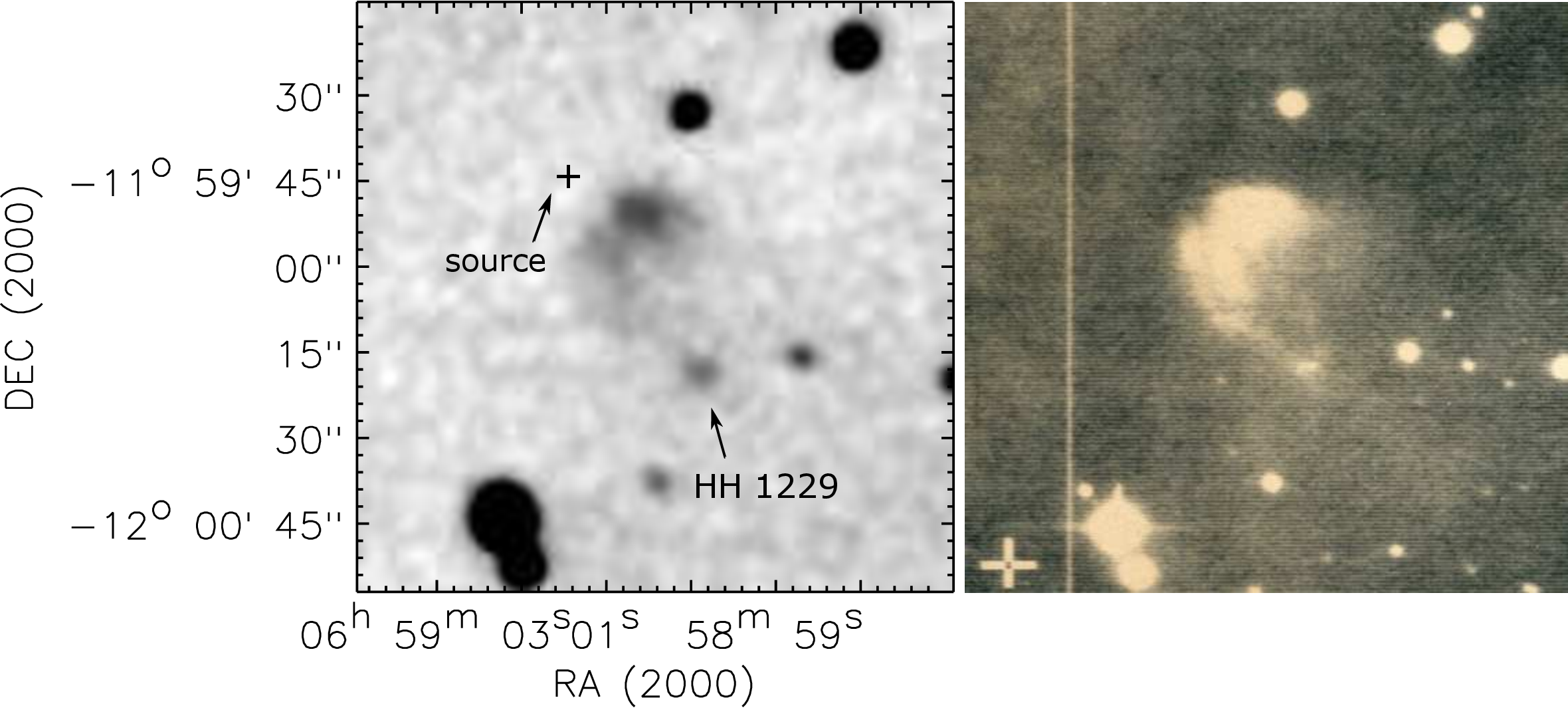}

   \caption{The RNO~80 nebula and the HH~1229 knot, visible on its axis.   1-m Schmidt, H$\alpha$ image (left panel); 3.6-m ESO telescope, broad-band R filter (right panel). The 2MASS 06590141$-$1159424 source inside the cloud is marked by a cross. }
   \label{RNO80}
\end{figure}

Our images (see Fig.\ref{RNO80}) reveal a single HH knot (HH~1229),
better visible in H$\alpha$ than in the [\ion{S}{ii}] lines, about
38\arcsec\ from the central star, and approximately on the symmetry
axis of the visible nebula. No other knots, nor signs of a
counterflow, have been detected. There are also no traces of an
emission jet inside the nebula. In the 4.6 $\mu$m image of the unWISE
survey at the place of HH~1229 one can see a faint patch, which
indicates the presence of H$_2$ emission.

The RNO 80 field is part of a large star-forming region CMa OB1 + CMa
R1. A compact group of stars with H$\alpha$ emission is located about
one degree to the west from the center of the main association
\citep{PR}, and this is where RNO~80 is located. On the basis of Gaia 
data the distance of this star forming region was estimated as 1185 pc \citep{PR}.
RNO 80 lies in a narrow dark lane that ends near the bright
reflection nebula GN 06.56.9.01. The illuminating star of this nebula
is Gaia DR3 3045899512500879744, the distance of which, according to
\citet{BJ2021}, is about 1330 pc.

\subsubsection{IRAS~19219+2300 field }

\begin{figure}
  \centering
  \includegraphics[width=\columnwidth]{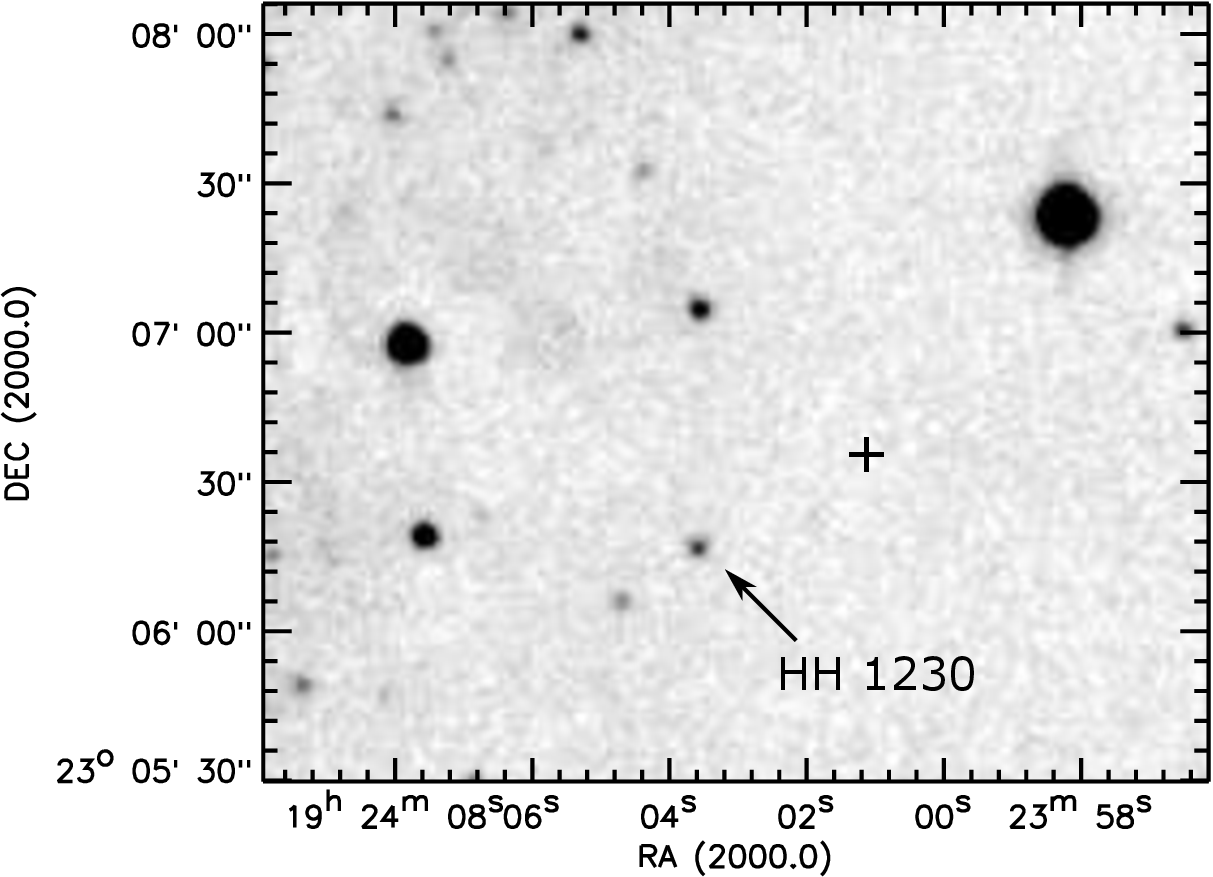}
   \caption{IRAS~19219+2300 (marked by a cross) and the HH~1230 knot. H$\alpha$ image.}
   \label{IRAS19219}
\end{figure}

This infrared source has no optical counterpart, and is located in the
dense dark cloud TGU~394P1 or Dobashi~1983 \citep{Dobashi2005,Dobashi2011}. The source is detected  in the 2MASS 
(as 19240107+2306351) and WISE surveys.  About 40\arcsec\ to the west
we have found a single compact HH knot (HH 1230) with nearly equal brightness in H$\alpha$ and [\ion{S}{ii}], for which the IRAS source seems to be the most probable source, since there are not any other suitable objects in the
vicinity (Fig.\ref{IRAS19219}). We have not found any references to
the distance of this cloud. However, on the western side of Dobashi
1983 two nebulous stars can be seen: Gaia DR3 2019298532316094208
and Gaia DR3 2022295079462085632, whose distances,
according to the \citet{BJ2021} catalog, are both around 530 pc. Thus,
we assume this value as the probable distance of the Dobashi 1983
cloud.


In the 4.6~$\mu$m IR image of the unWISE survey, the HH knot is
visible and has the appearance of a small bow shock, oriented towards
the east. We speculate that HH~1230 may
be part of a larger, perhaps partly embedded, flow.

\begin{figure}
  \centering
  \includegraphics[width=\columnwidth]{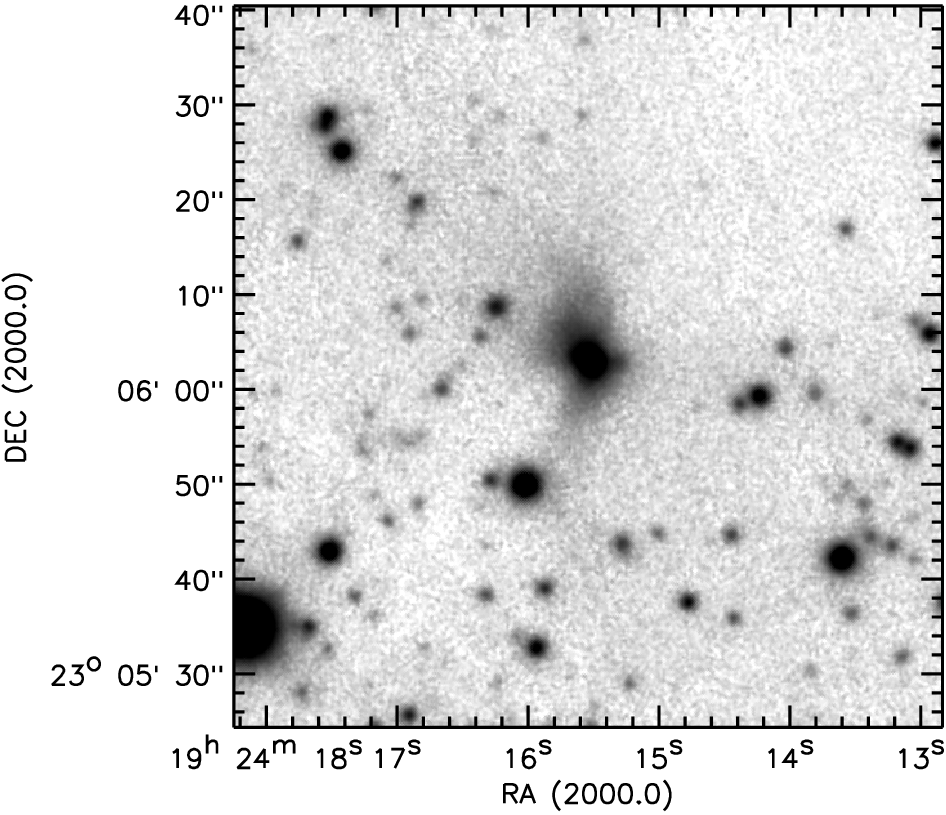}
   \caption{Newly found bipolar nebula. IRAS 19221+2300 coincides with a star in its center. PanSTARRS survey \textit{i} image.}
   \label{IRAS19221}
\end{figure}

We note that about 2.7\arcmin\ further to
the east from the HH~1230 knot we find another bright IR source, IRAS
19221+2300 (2MASS 19241551+2306030), which is also visible in the
optical. This star is located at the center of a so far unlisted
reflection nebula with a well-defined bipolar structure
(Fig. \ref{IRAS19221}). However, the axis of this nebula is nearly
orthogonal to the direction to HH~1230; thus it is unlikely to be the
exciting source of this object. The central star is listed as Gaia DR3
2019294237343467264. Its astrometric measurements are of very poor quality (RUWE = 5.45), but  it
appears likely that this object also belongs to the Dobashi~1983
cloud.  We have not found convincing evidence for any 
shock-excited emission in its environment on our wide-field
images.

\subsubsection{IRAS~20472+4338 field}

\begin{figure}
  \centering
  \includegraphics[width=\columnwidth]{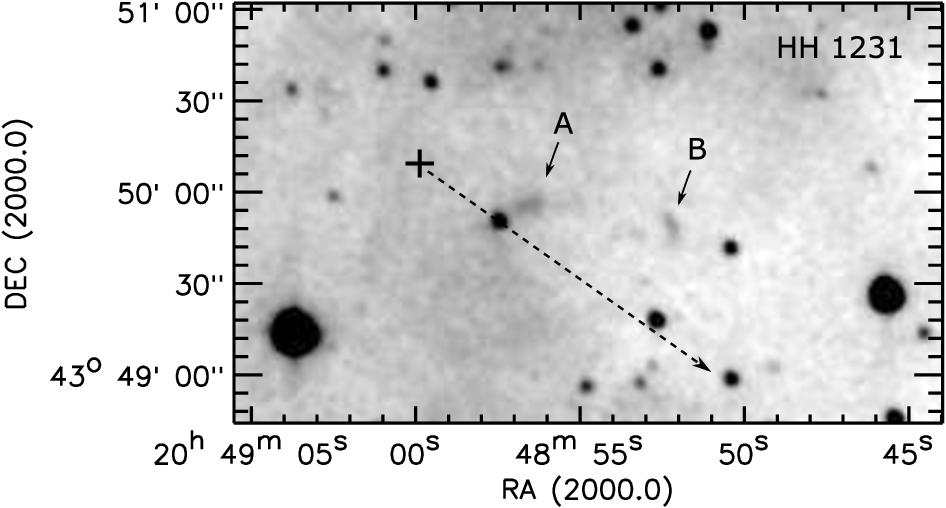}
   \caption{IRAS 20472+4338 (marked\ by cross) and two nearby HH knots (HH~1231 A and B), visible in  H$\alpha$ image. The approximate direction of the CO outflow from IRAS 20472+4338 is shown by dashed line.}  
   \label{IRAS_20472}
\end{figure}

This deeply embedded source is located within a dark cloud near the
western side of the Pelican nebula (IC~5070). It is not detected at
optical and near-IR J and H wavelengths and only appears weakly on a
2MASS \textit{K}-band of image, with  traces of nebulosity. This field  was covered by the Spitzer SEIP survey, where on the 3.6 and 4.5 $\mu$m images one can see a comma-like nebula, which extends $\sim15\arcsec$ 
from the IRAS
20472+4338 source in P.A. = 56\degr. Recently \citet{Zhang2020} found in $^{12}$CO evidence of a bipolar molecular outflow from this source \mbox{(DMOC-0006)}. The approximate direction of the red lobe of this outflow is shown on Fig.\ref{IRAS_20472}. Its P.A. (235\degr) shows that it is  
directed almost exactly opposite to the IR-nebula, seen near the source. 

This region is active in low mass star formation, and a number of HH
objects are located further north \citep{Bally2003}. We found two
faint HH knots (HH 1231 A and B) in the western direction from IRAS
20472+4338. Along with this source they define a nearly straight line
with P.A. = 255$\degr$ (Fig.\ref{IRAS_20472}). The HH knots probably lie within
the red-shifted lobe of the molecular outflow. Knot A is diffuse and
extended over about 10\arcsec. Knot B may have a bow-shape morphology,
which is more prominent in [SII].  We did not find evidence of H$_2$
emission in the images from allWISE and unWISE surveys. Their association to the IRAS
20472+4338 source seems possible but not entirely certain.

The distance of this molecular outflow as well as of the cloud is estimated by \citet{Zhang2020} as 900$\pm$100 pc. 

\section{Discussion and Conclusion}

The results presented in this paper, as well as data published
separately \citep{MMA,MMRA}, show that the survey technique employed
here is indeed successful, and new HH knots and flows have been found
near a significant part of the embedded IR sources observed. At least
two from the newly found HH flows (HH~1226 and HH~1227) lie in
isolated dark clouds, thus pointing to active star formation in these
regions. Other flows are also located in detached and dense globules
or filaments.

\begin{figure}
  \centering
  \includegraphics[width=\columnwidth]{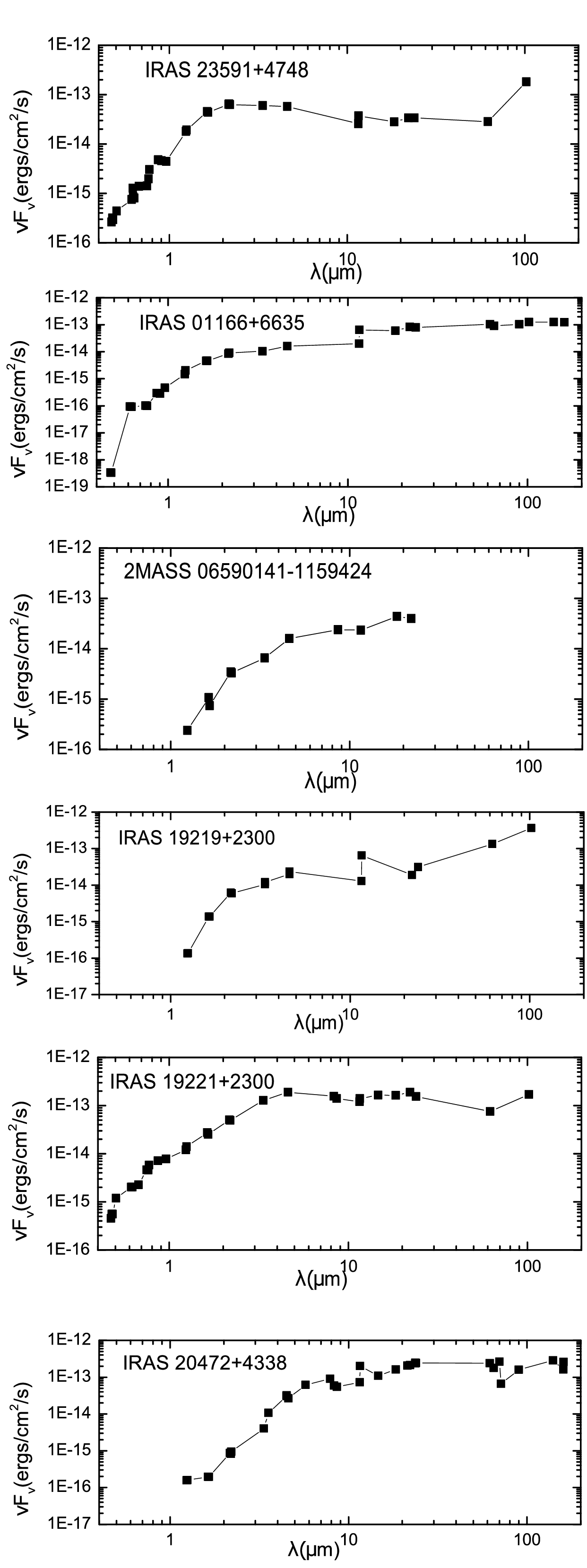} 

   \caption{SEDs of the infrared sources, associated with HH flows.}
   \label{SEDs_1}
\end{figure}

The IR sources which drive the newly discovered flows need further study.
With the aid of the \textit{VizieR Photometry viewer} we have built
spectral energy distributions (SEDs) for six IR sources from our
sample (Fig. \ref{SEDs_1}). IRAS 20472+4338 was also observed in two
wavebands of the Herschel PACS survey; we added these measurements to
its SED. The value of the 100~$\mu$m flux of this object, as measured
by IRAS, is improbably high, judging by the general appearance of its
SED; therefore, we excluded it from the final computations. We do not
include here the SED for WISEA J043041.15+352941.4, because it can be
found in the electronic version of the work of \citet{SST}. For
completeness we have also included the IRAS 19221+2300 source at the
center of the new bipolar nebula. One can see that all sources have
so-called "flat" SEDs, typical for very young objects, with the exception
of WISEA J043041.15+352941.4, which, according to \citet{SST}, has a
spectral index of about 1.5 with a rising SED, corresponding to a
Class I object.

As described in the previous sections, the distances of all objects
are now reasonably well known, which allows to estimate the bolometric
luminosity of the sources by integrating their SEDs.  The values thus
obtained are listed in Table \ref{lums}.  Of course, these values are
lower estimates, because extinction is not taken into account and, in addition, we do not have the data in the FIR/submm part of the SEDs. As
one can see, the objects have luminosities typical for low-mass YSOs, and
only the luminosity of IRAS 20472+4338 exceeds 10 $L_{\sun}$.

\begin{table}
\caption{Minimal bolometric luminosity of the infrared sources}       
\label{lums}     
\centering        
\begin{minipage}{\textwidth}
\begin{tabular}{l r c}          
\hline                        
 Source & D (pc) & $L (L_{\sun}$)  \\    
\hline                     
IRAS 23591+4748 (RNO 150) & 380  &  1.0   \\   
IRAS 01166+6635 & 240 &  0.6\\
WISEA J043041.15+352941.4 & 530 & 2.5$^{a}$ \\

2MASS 06590141$-$1159424 (RNO 80) & 1185 & 2.3 \\
IRAS 19219+2300 & 530 & 2.7 \\
IRAS 19221+2300 & 530 & 4.9  \\
IRAS 20472+4338 & 900 & 16.4 \\
\hline                                       
\end{tabular}

\footnotetext{$^{a}${From the survey of \citet{SST}.}}
\end{minipage}
\end{table}

\begin{figure}
  \centering
  \includegraphics[width=\columnwidth]{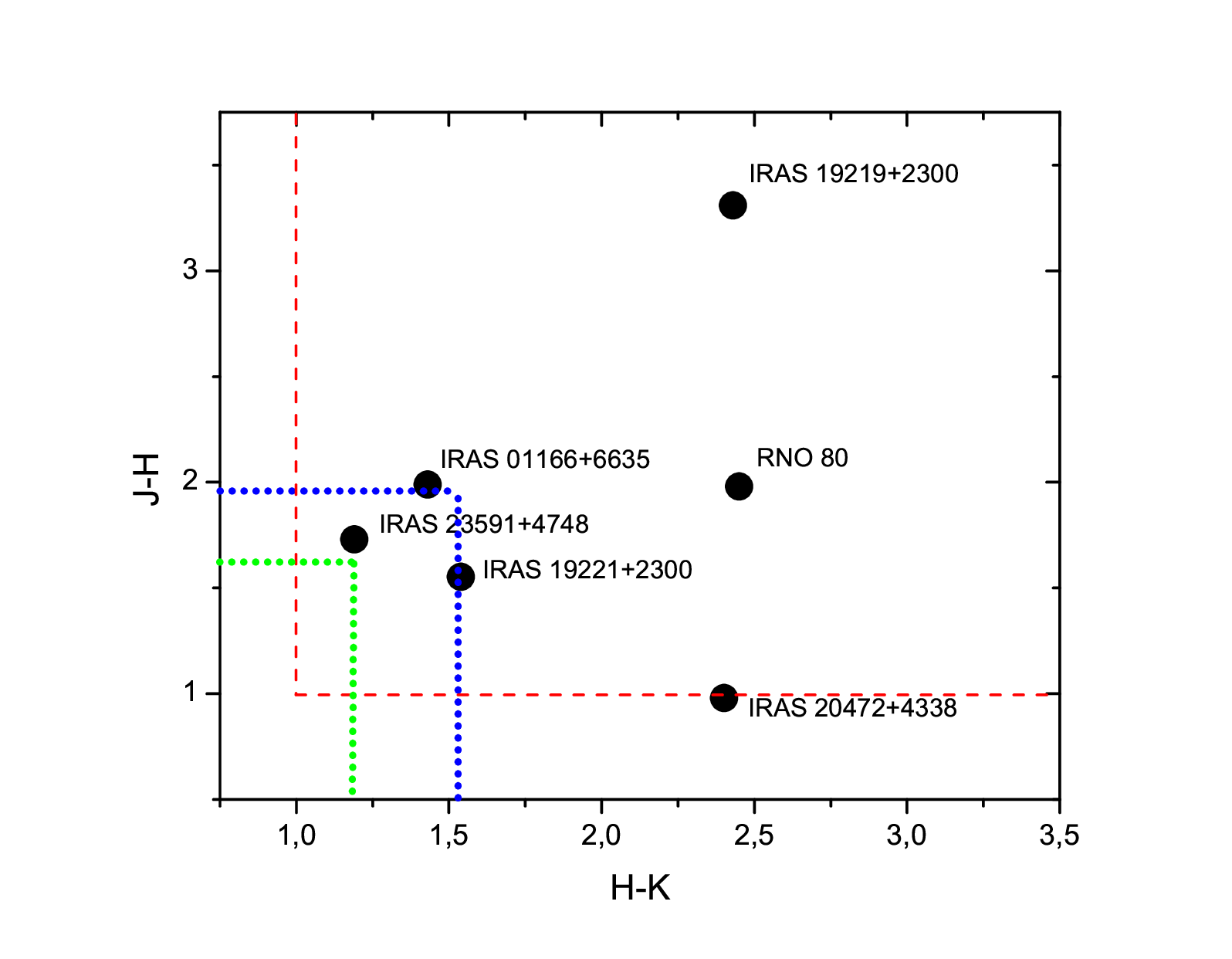}
   \caption{(J-H)/(H-K) diagram for the observed sources. Red rectangle shows the region of Class I sources, blue rectangle of HAeBe stars, and the green one of T Tau stars.}
   \label{diag}
\end{figure}

We have plotted all the IR sources discussed in this paper on a
(J-H)/(H-K) color-color diagram (Fig.\ref{diag}), with the exception
of WISEA J043041.15+352941.4, which was not detected in the near-IR
range. Also in this diagram the regions of Herbig Ae/Be and T Tau
stars and of the luminous Class I protostars, according to work of
\citet{Lopez}, are shown.  As can be seen in this figure, all of the
sources have IR color indices characteristic of YSOs, though they do
not correlate with their luminosity or with the shape of their SEDs
over the middle IR range. It is remarkable that the objects with $J-H$ and $H-K < 1$ are absent
in this sample. They all seem to be redder than the typical T Tau stars.

Three of the IR sources from our sample are visible in the optical range; their nature can be further investigated spectroscopically.
In fact, a program of long-slit spectroscopy of IR sources of HH flows, discernible also
in optical range, for the objects, found in Mon R1 association \citep{MMD} and in other fields, has already started, and some preliminary results were recently presented \citep{MM2023}.  

\section*{Acknowledgements}

We thank an anonymous referee for helpful comments, which improved the paper.

This work was supported by the RA MES State Committee of Science, in the frame of the research project number 21T-1C031.

This research has made extensive use of \textit{Aladin} sky atlas, \textit{VizieR}
catalogue access tool,  \textit{VizieR}
photometry tool and \textit{SIMBAD} database, which are developed and
operated at CDS, Strasbourg Observatory, France.  This work has made use of data from the European Space Agency (ESA) space mission Gaia, processed by the Gaia Data Processing and Analysis Consortium (DPAC). Funding for the DPAC is provided by national institutions, in particular the institutions participating in the Gaia MultiLateral Agreement (MLA). The Gaia mission website is \url{https://www.cosmos.esa.int/gaia}. The Gaia archive website is \url{https://archives.esac.esa.int/gaia}. This publication makes use of data products from the Two Micron All Sky Survey (2MASS), which is a joint project of the University of Massachusetts and the Infrared Processing and Analysis Center/California Institute of Technology, funded by the National Aeronautics and Space Administration and the National Science Foundation. This publication also makes use of data products from the Wide-field Infrared Survey Explorer (WISE), which is a joint project of the University of California, Los Angeles, and the Jet Propulsion Laboratory/California Institute of Technology, funded by the National Aeronautics and Space Administration. The images from Spitzer Enhanced Imaging Products program were used via NASA/IPAC
Infrared Science Archive \url{https://www.ipac.caltech.edu/doi/irsa/10.26131/IRSA3}

\section*{Data Availability}

The data underlying this article will be shared on reasonable request to the corresponding author.


\bsp    
\label{lastpage}
\end{document}